\documentclass[12pt]{article}
\usepackage{fullpage}
\usepackage{amsmath,graphicx}
\usepackage{amssymb} % \mathbb, etc.
\usepackage{bm}				% italic bold
\usepackage{cite}			% group citations
\usepackage{graphicx}		% figures
\usepackage{mathtools}		% for "coloneqq"
\usepackage{booktabs}
\usepackage{algorithm}
\usepackage{algpseudocode}
\usepackage{amsthm}			% theorems
\usepackage{enumitem}		% to have letter enumerations
\usepackage{multirow}       % to generate tables
\usepackage{tabularx}	    % to generate tables
\usepackage{hhline}
\usepackage{arydshln}		% dotted lines in a table
\usepackage{xcolor}
\usepackage{siunitx}

\newcolumntype{L}[1]{>{\raggedright\arraybackslash}p{#1}}
\newcolumntype{C}[1]{>{\centering\arraybackslash}p{#1}}
\newcolumntype{R}[1]{>{\raggedleft\arraybackslash}p{#1}}

\def\defn{\,\triangleq\,} % definitions: equality with triangle
\def\d{\, \mathrm{d}} % differential
\def\Ein{E_\mathsf{in}}
\def\Esc{E_\mathsf{sc}}

\def\cbf{\mathbf{c}}
\def\ebf{\mathbf{e}}

\def\fbf{\mathbf{f}}

\def\Sbf{\mathbf{S}}

\def\Jbf{\mathbf{J}}
\def\Ebf{\mathbf{E}}
\def\Ebfin{\Ebf_\mathsf{in}}
\def\Ebfsc{\Ebf_\mathsf{sc}}
\def\Gbf{\mathbf{G}}

\def\rbm{\bm{r}}

\def\rbmp{{\bm{r}^\prime}}

\def\Lcal{\mathcal{L}}
\def\Scal{\mathcal{S}}

\def\Mcal{\mathcal{M}}

\def\R{\mathbb{R}}

\begin{document}

\title{Neural Inverse Scattering with Score-based Regularization}

\author{
Yuan~Gao$^{\footnotesize 1}$, Wenhan Guo$^{\footnotesize 2}$,~and~Yu Sun$^{\footnotesize 1, \ast}$ \\
\emph{\footnotesize $^{\footnotesize 1}$Johns Hopkins University, MD 21218, USA}\quad\emph{\footnotesize $^{\footnotesize 2}$Pomona College, CA 91711, USA}\\
\small$^{\footnotesize *}$\emph{Email}: \texttt{ysun214@jh.edu}
}
\maketitle
\begin{abstract}
Inverse scattering is a fundamental challenge in many imaging applications, ranging from microscopy to remote sensing.
Solving this problem often requires jointly estimating two unknowns---the image and the scattering field inside the object---necessitating effective image prior to regularize the inference.
In this paper, we propose a regularized \textit{neural field (NF)} approach which integrates the \textit{denoising score function} used in score-based generative models.
The neural field formulation offers convenient flexibility to performing joint estimation, while the denoising score function imposes the rich structural prior of images.
Our results on three high-contrast simulated objects show that the proposed approach yields a better imaging quality compared to the state-of-the-art NF approach~\cite{Luo.etal2024}, where regularization is based on total variation. 
\end{abstract}
\section{Introduction}
\label{sec:intro}
The inverse scattering problem aims to reconstruct an object's dielectric permittivity from the measurements of its scattered electromagnetic field. 
It arises in various imaging applications, including tomographic microscopy~\cite{Choi.etal2007}, digital holography~\cite{Brady.etal2009, Pellizzari.etal2020}, and radar imaging~\cite{Liu.etal2016a}. 
Nonlinear scattering models are essential in scenarios involving strong scattering, as they account for multiple scattering interactions within the object~\cite{Belkebir.Sentenac2003, Belkebir.etal2005}. Recent studies have explored methods that either predefine the nonlinear scattering model under specific conditions~\cite{Soubies.etal2017, Chowdhury.etal2019} or jointly estimate it during reconstruction~\cite{Liu.etal2018, Ma.etal2018}.

\begin{figure*}[t]
\centering
\includegraphics[width=0.8\textwidth]{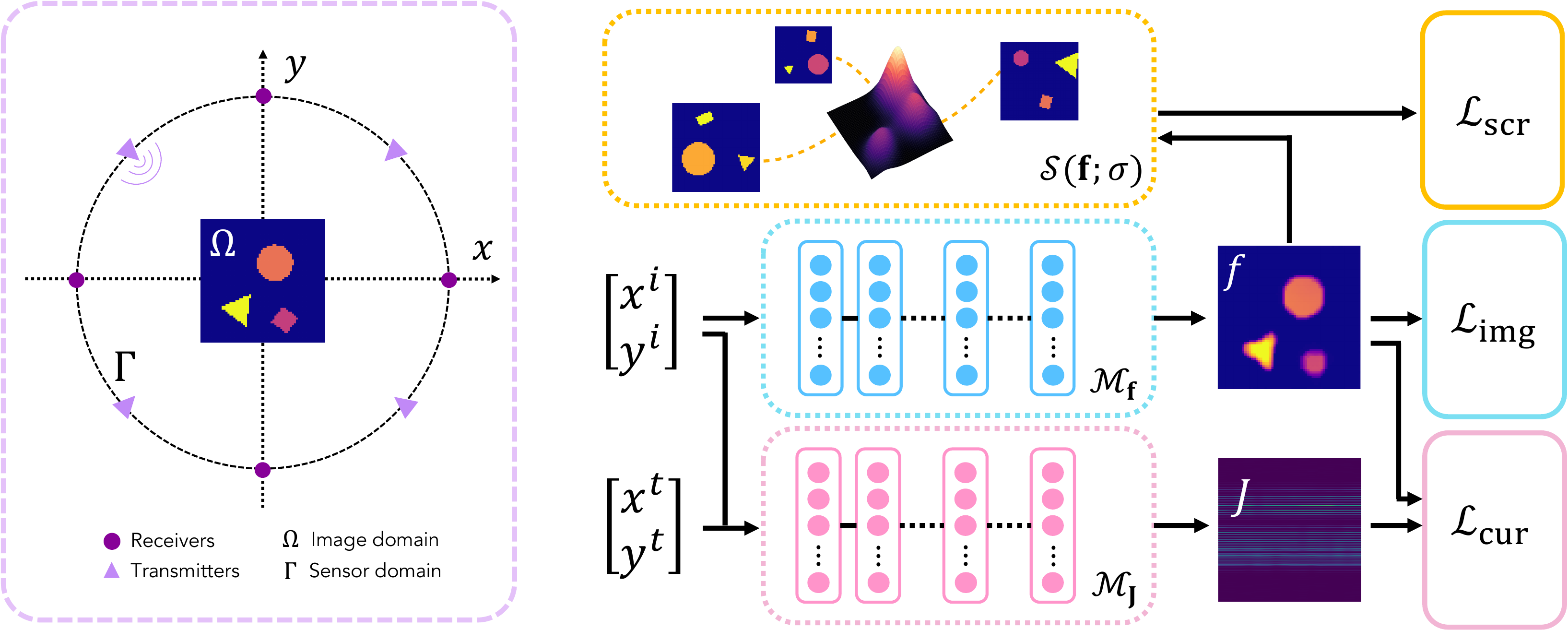}
\caption{\textit{Left:} Visualization of the experimental setup for the inverse scattering problem under consideration. The object within the domain $\Omega$ is illuminated by transmitters and observed by receivers, both distributed along a circular sensor orbit $\Gamma$. \textit{Right:} Conceptual illustration of the proposed \textit{NF-Score} method, which employs two MLPs, $\mathcal{M}_f$ and $\mathcal{M}_J$, to estimate $\fbf$ and $\Jbf$, respectively. A score-based regularization term is added into the loss function to enhance reconstruction quality.}
\label{Fig:Scheme}
\end{figure*}

\textit{Neural fields (NFs)} have recently emerged as a powerful deep learning paradigm in computer graphics~\cite{Mildenhall.etal2020} and computational imaging~\cite{Sun.etal2022}. 
Unlike traditional end-to-end deep learning models~\cite{Sun.etal2018}, NFs learn a coordinate-based mapping from input coordinates to target quantities (e.g., pixel values) by minimizing their discrepancy from collected measurements. 
This flexible formulation enables NFs to infer multiple unknown quantities simultaneously. For example, they have been used to jointly estimate the scene and aberration error in wavefront shaping~\cite{Feng.etal2023neuws} and object structure along with motion in computed tomography~\cite{Reed.etal2021}. More recently, Luo et al.~\cite{Luo.etal2024} applied NFs to inverse scattering, jointly estimating intra-object scattering and dielectric permittivity. However, such joint inference problems are inherently non-convex and require effective regularization.

In this work, we propose a regularized NF-based inverse scattering solver by integrating the \textit{denoising score function}, which is trained to denoise the additive white Gaussian noise (AGWN). 
In the field of score-based diffusion models, it has been shown that such a denoiser approximates the \textit{score} (i.e. gradient of log-density) of the image distribution~\cite{Song.etal2021}. 
By including the $\ell_2$ norm of the denoising score function into the loss function, our method produces a solution that better captures the image statistics. We validate our method, named \textit{NF-Score}, on three test objects, demonstrating both visually (Fig.\ref{Fig:Scheme}) and numerically (Tab.\ref{Tab:Numerical}) that it outperforms the state-of-the-art NF-based method paired with total variation~\cite{Luo.etal2024}.

\section{Inverse Scattering Setup}
We consider the problem of reconstructing an object's permittivity distribution from measurements of scattered electromagnetic waves. As shown in Fig.~\ref{Fig:Scheme} (\textit{Left}), an object of the permittivity distribution $\epsilon(\rbm)$ is confined within a bounded sample domain ($\Omega \subseteq \R^2$) and immersed in a background medium with permittivity $\epsilon_b$. 
Following the 2D experimental setup in~\cite{Wei.etal2018, Luo.etal2024}, we arrange the transmitters and receivers along a circular sensor orbit ($\Gamma\subseteq\R^2$) with the object placed in the center. In total, we use $n_t=16$ transmitters and $32$ receivers, where each transmitter sequentially illuminates the object with a known incident field $E_\mathsf{in}$. The electromagnetic interaction between the incident field and the object generates scattered fields $E_\mathsf{sc}$ that are measured by the receivers. 

For a single illumination $\Ein$, the relationship between the object and the electromagnetic field in the sample plane is governed by the Lippmann-Schwinger equation~\cite{Born.Wolf2003}
\begin{equation}
\label{Eq:ImageField}
E(\rbm) = E_\mathsf{in}(\rbm) + k^2\int_{\Omega} g(\rbm - \rbmp) \, \underbrace{f(\rbmp) \, E(\rbmp)}_{\triangleq J(\rbm)} \d \rbmp, \;(\rbm\in\Omega)    
\end{equation}
where $E(\rbm)$ is the total wave field, $k = 2\pi/\lambda$ the wavenumber, and $f(\rbm) \defn \epsilon(\rbm)-\epsilon_b$ the permittivity contrast. For convenience, it is also common to introduce the induced current, defined as $J(\rbm) \defn f(\rbm)E(\rbm)$~\cite{Luo.etal2024}.
In 2D free space, Green's function $g(\rbm)$ is defined as
\begin{equation}
\label{Eq:GreenFunction}
g(\rbm) \triangleq \frac{j}{4} H_0^{(1)}\left(k_b\|\rbm\|_{2}\right).
\end{equation}
Here, $k_b \triangleq k \sqrt{\epsilon_b}$ is the wavenumber of the background medium, and $H_0^{(1)}$ is the zero-order Hankel function of the first kind.
The scattered light field measured at the sensor plane is given by 
\begin{equation}
\label{Eq:ScatteredField}
E_{\mathsf{sc}}(\rbm)=k^2\int_{\Omega} g\left(\rbm-\rbm^{\prime}\right) f\left(\rbm^{\prime}\right) E\left(\rbm^{\prime}\right) d \rbm^{\prime}, \;(\rbm\in\Gamma).
\end{equation}
The discrete system that models the full wave-object interaction is given by
\begin{subequations}
\label{Eq:NonlinearModel}
\begin{align}
&\Ebf = \Ebfin + \Gbf \cdot\Jbf \label{Eq:ForwardScat2} \\
&\Ebfsc = \Sbf\cdot (\fbf \odot \Ebf) + \ebf \label{Eq:ForwardScat1}\;, 
\end{align}
\end{subequations}
Here, $\Ebf$, $\Ebfin$, $\Ebfsc$, and $\Jbf$ are the discrete representations of $E$, $\Ein$, $\Esc$, and $J$, respectively, and $\ebf$ accounts for measurement noise. The induced current is expressed as $\mathbf{J} \triangleq \mathbf{f} \odot \mathbf{E}$, where $\odot$ denotes the element-wise product. Additionally, $\Gbf$ is the discretized Green's function within the sample plane, while $\Sbf$ is the discretized Green's function mapping from $\Omega$ to $\Gamma$. 

\begin{figure*}[t!]
\centering
\includegraphics[width=0.97\textwidth]{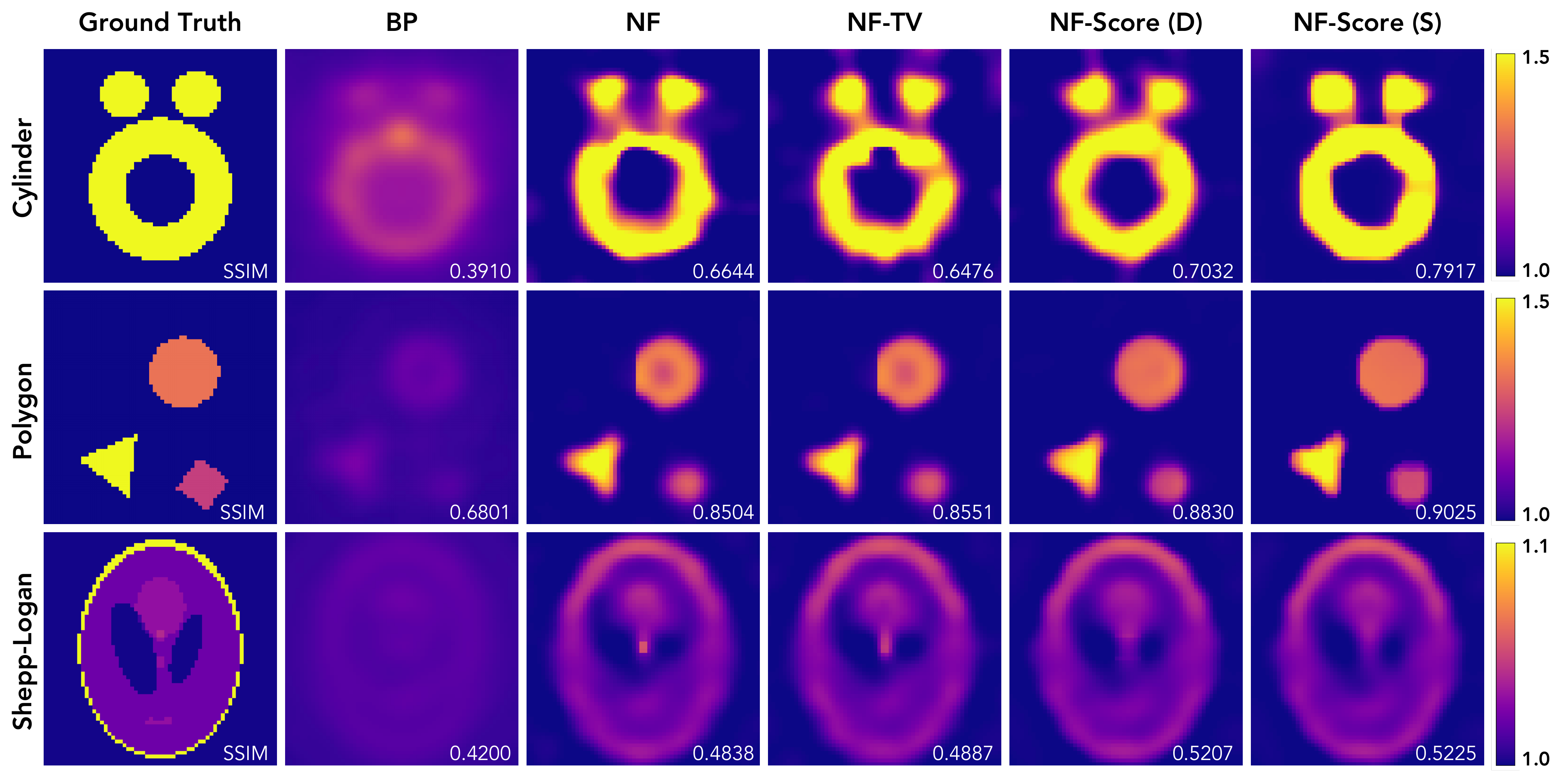}
\caption{Visual comparison of the permittivity contrast images reconstructed by the proposed NF-Score method and baseline methods. We present two variants: \textit{NF-Score (D)}, which utilizes a pre-trained DnCNN denoiser, and \textit{NF-Score (S)}, which incorporates a pre-trained score network. Both variants demonstrate superior reconstruction quality and reduced artifacts compared to the baselines, which is also corroborated by the SSIM values attached to each image.}
\label{Fig:Visual}
\end{figure*}

\begin{table*}[t!]
    \centering
    \caption{The SSIM and PSNR values obtained by NF-Score and baseline methods on test objects \textit{Au}, \textit{Polygon}, and \textit{Shepp-Logan}. Best values are highlighted in \textbf{bold}.}
    \label{Tab:Numerical}
    \small
    \vspace{0.5ex}
    \begin{tabular*}{0.98\textwidth}{C{80pt}C{50pt}C{50pt}C{50pt}C{50pt}C{50pt}C{50pt}}
        \toprule
        \multirow{2}{*}{\textbf{Methods}} & \multicolumn{2}{c}{\textbf{Au}} & \multicolumn{2}{c}{\textbf{Polygon}} & \multicolumn{2}{c}{\textbf{Shepp-Logan}} \\
        \cmidrule(lr){2-3} \cmidrule(lr){4-5} \cmidrule(lr){6-7}
        & SSIM & PSNR & SSIM & PSNR & SSIM & PSNR \\ [0.1ex]
        \midrule \\ [-2.1ex]
        BP & $0.3910$ & $18.43$ & $0.6801$ & $23.29$ & $0.4200$ & $34.94$ \\ [0.3ex]
        NF & $0.6644$ & $22.30$ & $0.8504$ & $31.35$ & $0.4838$ & $36.76$ \\ [0.3ex]
        NF-TV~\cite{Luo.etal2024} & $0.6476$ & $23.06$ & $0.8551$ & $31.44$ & $0.4887$ & $36.78$ \\ [0.3ex]
        \hdashline[0.7pt/2pt]
        \noalign{\vskip 1ex}
        NF-Score (D) & $0.7032$ & $24.28$ & $0.8830$ & $32.17$ & $0.5207$ & $36.97$ \\ [0.3ex]
        NF-Score (S) & $\mathbf{0.7917}$ & $\mathbf{24.93}$ & $\mathbf{0.9025}$ & $\mathbf{32.35}$ & $\mathbf{0.5225}$ & $\mathbf{37.03}$ \\ [0.1ex]
        \bottomrule
    \end{tabular*}
\end{table*}

\section{Proposed Method}

Fig.~\ref{Fig:Scheme} (\textit{Right}) provides a conceptual illustration of the proposed method, which consists of two multi-layer perceptrons (MLPs) to estimate $\fbf$ and $\Jbf$, respectively. The first MLP, $\mathcal{M}_f$, takes the spatial coordinates $\cbf_\Omega=(x^i, y^i)$ as input and estimates the scattering potential $\mathbf{f}$, while the second MLP, $\mathcal{M}_J$, additionally takes the transmitter positions $\cbf_\Gamma = (x^t, y^t)$ as input to estimate the induced current $\mathbf{J}$. We note that this architecture follows the one proposed in~\cite{Luo.etal2024}.
The loss functions for learning these two quantities are
\begin{subequations}
\begin{align}
&\Lcal_\mathsf{img} = \sum_{n_t}\| \Ebfsc - \Sbf\cdot (\hat{\fbf} \odot \tilde{\Ebf}) \|_2^2, &\quad \hat{\fbf} = \Mcal_\fbf(\cbf_\Omega)  \\
&\Lcal_\mathsf{cur} = \sum_{n_t}\| \tilde{\Jbf} - \hat{\Jbf} \|_2^2, &\quad \hat{\Jbf} = \Mcal_\Jbf(\cbf_\Gamma)
\end{align}
\end{subequations}
where $\tilde{\Ebf}$ is the estimated total field computed by using the predicted $\hat{\Jbf}$ according to~(\ref{Eq:NonlinearModel}a), and $\tilde{\Jbf}$ is the recomputed induced current using $\tilde{\Jbf} = \fbf\odot\Ebfin + \fbf\odot(\Gbf\cdot\Jbf)$. 
Note that $\Lcal_\mathsf{cur}$ imposes the self-consistency for $\Jbf$ according to the physics of wave propagation.

To regularize the training process, we propose to include the following term in the final loss function
\begin{equation}
\Lcal_\mathsf{scr} = \frac{1}{2}\|\Scal(\fbf;\sigma)\|_2^2
\end{equation}
where $\Scal(\fbf;\sigma)$ is the learned denoising score function with $\sigma$ controls its denoising strength. Note that $\Scal(\cdot;\sigma)$ is pre-trained, and its parameters are not updated during the training. Mathematically, the denoising score function approximates $\Scal(\fbf;\sigma) \approx \nabla \log p_\sigma(\fbf)$, where $p_\sigma(\fbf)$ is distribution of the noisy image $\fbf + \text{AWGN}$; yet, when $\sigma$ is small, $p_\sigma(\fbf) \approx p(\fbf)$ which is the distribution of clean images. The final training loss is given by the weighted sum of all loss terms
\begin{equation}
\Lcal_\mathsf{train} = \alpha_1 \Lcal_\mathsf{img} + \alpha_2 \Lcal_\mathsf{cur} + \alpha_3 \Lcal_\mathsf{scr}.
\end{equation}
For all experiments, we trained the proposed model until convergence by monitoring the value of the total loss.

\begin{figure}[t!]
\centering
\includegraphics[width=0.45\textwidth]{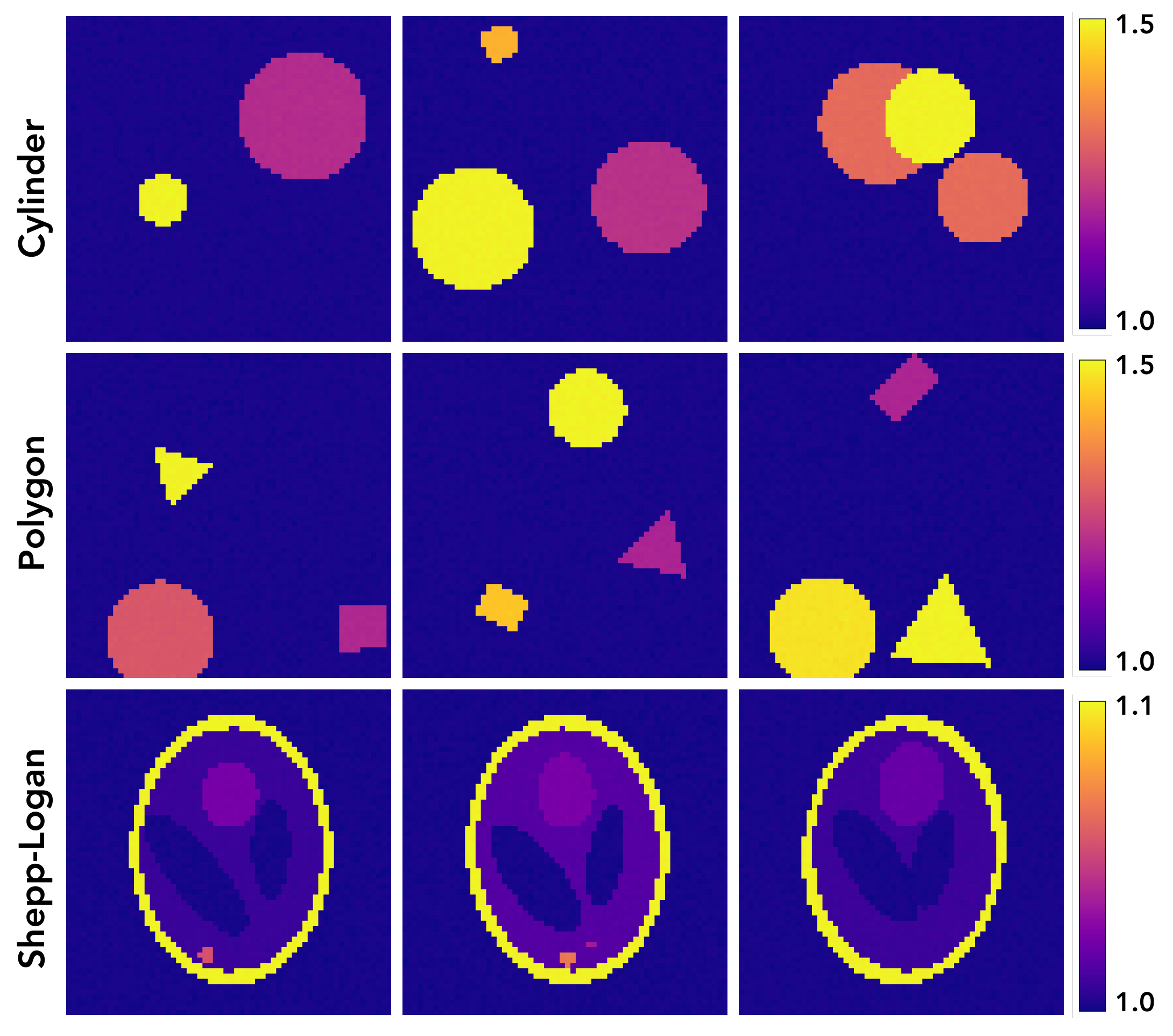}
\caption{Visualization of the samples generated by running annealed Langevin dynamics~\cite{Song.etal2019} equipped with the score networks trained on the synthetic datasets corresponding to \textit{Au}, \textit{Polygon}, and \textit{Shepp-Logan}, respectively.}
\label{Fig:score_generation}
\end{figure}

\begin{figure}[t!]
\centering
\includegraphics[width=0.45\textwidth]{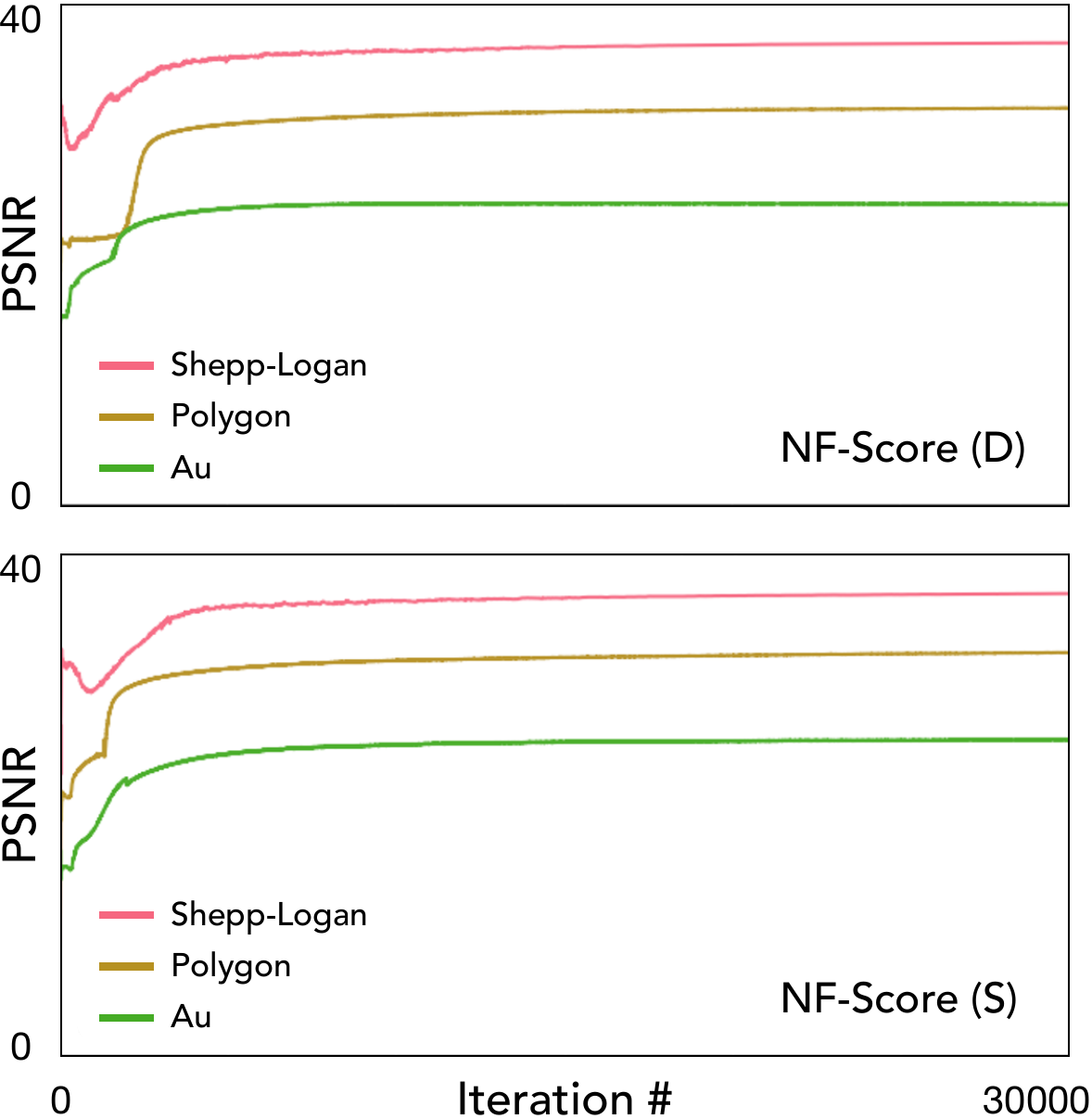}
\caption{Illustration of the convergence of NF-Score (D) and NF-Score (S) during training. The PSNR value are plotted against the iteration number.}
\label{Fig:convergence}
\end{figure}

\begin{figure}[t!]
\centering
\includegraphics[width=0.45\textwidth]{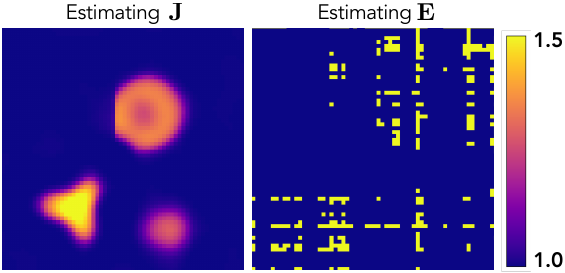}
\caption{Comparison of the reconstructed permittivity contrast obtained by estimating $\Jbf$ versus $\Ebf$. 
NF was selected as the reconstruction method to eliminate the influence of regularization.
}
\label{Fig:ablation}
\end{figure}

\section{Experimental Results}

\textbf{Experimental setup}\quad
We validate NF-Score on three test objects: \textit{Au}, \textit{Polygon} and \textit{Shepp-Logan}. We generate the measurements by performing the full wave propagation described in~\cite{Wei.etal2018}, further corrupted by 30\% (\textit{Au}) and 5\% (\textit{Polygon} and \textit{Shepp-Logan}) AWGN, respectively. We implement two variants of our regularization approach: \textit{NF-Score (D)}, which approximates the score function $\Scal(\cdot;\sigma)$ using the pre-trained denoiser ($\sigma=7/255$) in~\cite{Liu.2021}, and \textit{NF-Score (S)}, which directly employs a score network.
We adopt the strategy in~\cite{Sun.etal2024} for training the networks. 
We construct three training datasets that preserve the key geometrical structures of the test objects while introducing randomized locations and contrasts.
Fig.~\ref{Fig:score_generation} visualizes the samples generated by each score network using the annealed Langevin dynamics~\cite{Song.etal2019}.

We compute \textit{structural similarity index measure (SSIM)} and \textit{peak signal-to-noise ratio (PSNR)} to evaluate reconstruction quality and consider backpropagation (\textit{BP}), vanilla NF (\textit{NF}), and NF regularized by TV (\textit{NF-TV})~\cite{Luo.etal2024} as baselines. We set $\alpha_1 = \alpha_2 = 1$ for all NF-based methods. As for $\alpha_3$, we adopt the recommended value of $\alpha_3 = 0.01$ from the prior work for NF-TV~\cite{Luo.etal2024}, while for our score-based approaches, we finetune $\alpha_3$ to optimize performance for each object.

\smallskip
\noindent
\textbf{Results}\quad
Fig.~\ref{Fig:Visual} and Tab.~\ref{Tab:Numerical} present the visual and quantitative comparisons across all three test objects. As shown, the BP method fails to recover meaningful structures, while the NF approach yields improved reconstructions but still suffers from visible artifacts, even with TV regularization. 
In contrast, both variants of the proposed NF-Score method substantially reduce artifacts and enhance image quality. 
These gains are numerically validated in Tab.\ref{Tab:Numerical}. NF-Score (D) consistently outperforms NF-TV, achieving SSIM improvements of at least 0.025 across all test objects and a PSNR gain of 1.22 dB for the \textit{Au} target. 
NF-Score (S) achieves the highest performance across all metrics, with SSIM gains of 0.144 for \textit{Au}, 0.047 for \textit{Polygon}, and 0.034 for \textit{Shepp-Logan}, and corresponding PSNR improvements of 1.87 dB, 0.91 dB, and 0.25 dB, respectively. 
These improvements highlight the effectiveness of score-based regularization. 
Furthermore, as illustrated in Fig.~\ref{Fig:convergence}, the proposed method exhibits stable optimization dynamics, with reconstruction quality improving steadily during training.

Lastly, we justify our choice of estimating the induced current $\Jbf$ rather than the total light field $\Ebf$. 
Fig.~\ref{Fig:ablation} compares the reconstructed permittivity contrast obtained by estimating $\Jbf$ versus $\Ebf$. 
We used NF as the reconstruction method to eliminate the influence of regularization. 
It is clear that directly estimating $\Ebf$ leads to a failure in the reconstruction, caused by the divergence in MLP training.

\section{Conclusion}
We introduced a NF-based inverse scattering solver with score-based regularization to enhance image reconstruction. By integrating a pre-trained denoising score function, our method effectively enforces expressive image priors, reducing artifacts and improving accuracy. Experimental results on synthetic datasets show superior performance over TV. Future work will explore the robustness of our approach under different noise models beyond AWGN.

\end{document}